%% file: pyrochlore.tex
\documentclass[aps,prl,amsmath,amssymb,twocolumn,superscriptaddress,floatfix]{revtex4-1}
\usepackage{graphicx}
\usepackage{hyperref}
\hypersetup{pdftitle={Pyrochlore Mn2Sb2O7},pdfauthor={Darren C. Peets},pdfsubject={Frustrated magnetism},pdfdisplaydoctitle}

\def\Mn{Mn\ensuremath{_2}Sb\ensuremath{_2}O\ensuremath{_7}}
\def\py{pyr-Mn\ensuremath{_2}Sb\ensuremath{_2}O\ensuremath{_7}}

\def\TSG{\ensuremath{T_\text{f}}}
\def\TCW{\ensuremath{T_\text{CW}}}
\def\P3{\ensuremath{P3_121}}
\begin{document}

\title{The $3d$-Electron Heisenberg Pyrochlore Mn$_2$Sb$_2$O$_7$}

\author{Darren C. Peets}
\altaffiliation[Current address: ]{State Key Laboratory of Surface Physics, Department of Physics, and Advanced Materials Laboratory, Fudan University, Shanghai 200438, China}
\author{Hasung Sim}
\affiliation{Center for Correlated Electron Systems, Institute for Basic Science (IBS), Seoul 08826, Korea}
\affiliation{Department of Physics and Astronomy, Seoul National University, Seoul 08826, Korea}

\author{Maxim Avdeev}
\affiliation{Australian Nuclear Science and Technology Organisation, Lucas Heights, NSW 2234, Australia}

\author{Je-Geun Park}
\affiliation{Center for Correlated Electron Systems, Institute for Basic Science (IBS), Seoul 08826, Korea}
\affiliation{Department of Physics and Astronomy, Seoul National University, Seoul 08826, Korea}


\begin{abstract}

  In frustrated magnetic systems, geometric constraints or the
  competition amongst interactions introduce extremely high degeneracy
  and prevent the system from readily selecting a low-temperature
  ground state.  The most frustrated known spin arrangement is on the
  pyrochlore lattice, but nearly all magnetic pyrochlores have
  unquenched orbital angular momentum, constraining the spin
  directions through spin-orbit coupling.  Pyrochlore \Mn\ is an
  extremely rare Heisenberg pyrochlore system, with
  directionally-unconstrained spins and low chemical disorder.  We
  show that it undergoes a spin-glass transition at 5.5\,K, which is
  suppressed by disorder arising from Mn vacancies, indicating this
  ground state to be a direct consequence of the spins' interactions.
  The striking similarities to $3d$ transition metal pyrochlores with
  unquenched angular momentum suggests that the low spin-orbit
  coupling in the $3d$ block makes Heisenberg pyrochlores far more
  accessible than previously imagined.

\end{abstract}

\maketitle




Strong frustration, in which interactions compete, impedes spin
systems from selecting a unique global ground state at low
temperature, leading to a wide variety of physics in which
fluctuations, quantum mechanical effects, and fine details of the
spin-spin interactions can be crucial\cite{Lacroix2011}.  Due to their
extremely strong magnetic frustration, pyrochlore oxides and halides
host a plethora of exotic phases, such as quantum spin
liquids\cite{Gardner2010,Balents2010,Ross2011}, or emergent magnetic
monopoles\cite{Castelnovo2008,Morris2009,Fennell2009} for classical
spins.  However, the spins in nearly every known magnetic pyrochlore
are either Ising- or XY-like, constrained through spin-orbit coupling
to point directly into or out of the tetrahedra on whose corners they
reside, or in a plane perpendicular, and in no other direction.
Pyrochlore lattices with directionally-unconstrained Heisenberg spins
are scarce and not as well studied.  With far more degrees of freedom
leading to far greater degeneracy, but also a greater ability to adapt
to interactions, Heisenberg pyrochlore systems may offer immense
potential for unveiling new physics\cite{Wan2016}.

Spin-orbit coupling, the interaction responsible for coupling spins to
the crystal lattice thereby constraining their directions, strengthens
as the atomic number increases.  Heavy magnetic ions, such as the
lanthanides most commonly encountered in the pyrochlore structure, are
thus best considered in terms of spin-orbit-coupled $j$-states linked
to the lattice.  A comparative lack of pyrochlores containing magnetic
ions from earlier in the periodic table, particularly the $3d$ block,
makes pure Heisenberg physics scarce in this lattice.  The
best-studied $3d$-electron pyrochlores, spinel oxides in which the
B-site forms a pyrochlore lattice, typically contain Cr$^{3+}$
($s=3/2$)\cite{Lee2002,Tomiyasu2008}, whose orbital moments should not
be fully quenched and may not behave as pure Heisenberg spins.  To the
authors' knowledge, only three known materials host pure Heisenberg
spins in a pyrochlore lattice: FeF$_3$\cite{dePape1986,Reimers1991},
the pyrochlore variant of
\Mn\ (\py)\cite{Brisse1972,Zhou2008,Zhou2010}, and recently reported
NaSrMn$_2$F$_7$\cite{Sanders2016}.  All have a high-spin $3d^5$
electron configuration with fully quenched orbital moments, and both
FeF$_3$ and \py\ are challenging to prepare by conventional
solid-state synthesis.  Pyrochlore FeF$_3$ orders magnetically at
22\,K\cite{dePape1986,Calage1987}, a more than 10-fold suppression
relative to its less-symmetric and less-frustrated rhombohedral
polymorph\cite{Ferey1986}, while the 2.7\,K transition in
NaSrMn$_2$F$_7$ also indicates very strong frustration.  Curiously,
\py\ has been reported to form a spin glass around
41\,K\cite{Zhou2008,Zhou2010}, several times {\slshape higher} than
the magnetic transition in its less-frustrated
\P3\ polymorph\cite{Reimers1991,Peets-P2}\footnote{The \P3\ polymorph
  of \Mn\ consists of alternating kagome and triangular Mn layers, as
  in the pyrochlore structure, but shifted to produce a very
  different, chiral, structure.  A detailed study on the
  \P3\ polymorph is presented elsewhere\cite{Peets-P2}}.  With
\py\ having a Curie-Weiss temperature just under 50\,K, its putative
41\,K transition would correspond to remarkably low frustration in the
most-frustrated known three-dimensional magnetic lattice.


In a Heisenberg pyrochlore, the most likely magnetic ground state
would be a counterpart of the all-in-all-out state found in Ising
pyrochlore antiferromagnets, but the considerable degeneracy could
make a variety of other states possible\cite{Wan2016}, or may prevent
long-range order\cite{Moessner1998PRL,Moessner1998PRB}.  The lack of
purely-Heisenberg systems and the anomalously low frustration in
\py\ have hampered the search for novel physics. We show that the spin
glass transition in \py\ is actually 5.5\,K, corresponding to much
higher frustration, and this transition is suppressed by Mn vacancies.
The Mn$^{2+}$ magnetic ion adopts a $3d^5$ high-spin electronic
configuration, with no orbital degree of freedom, leaving the spin
directions fully unconstrained.  The magnetic behavior is strikingly
similar to that in several closely-related $3d$ pyrochlores having
unquenched orbital moments, indicating that the low spin-orbit
coupling in the $3d$ transition metals may render a much broader
family of materials as effective Heisenberg pyrochlore systems.

\begin{figure}[htb]
  \includegraphics[width=\columnwidth]{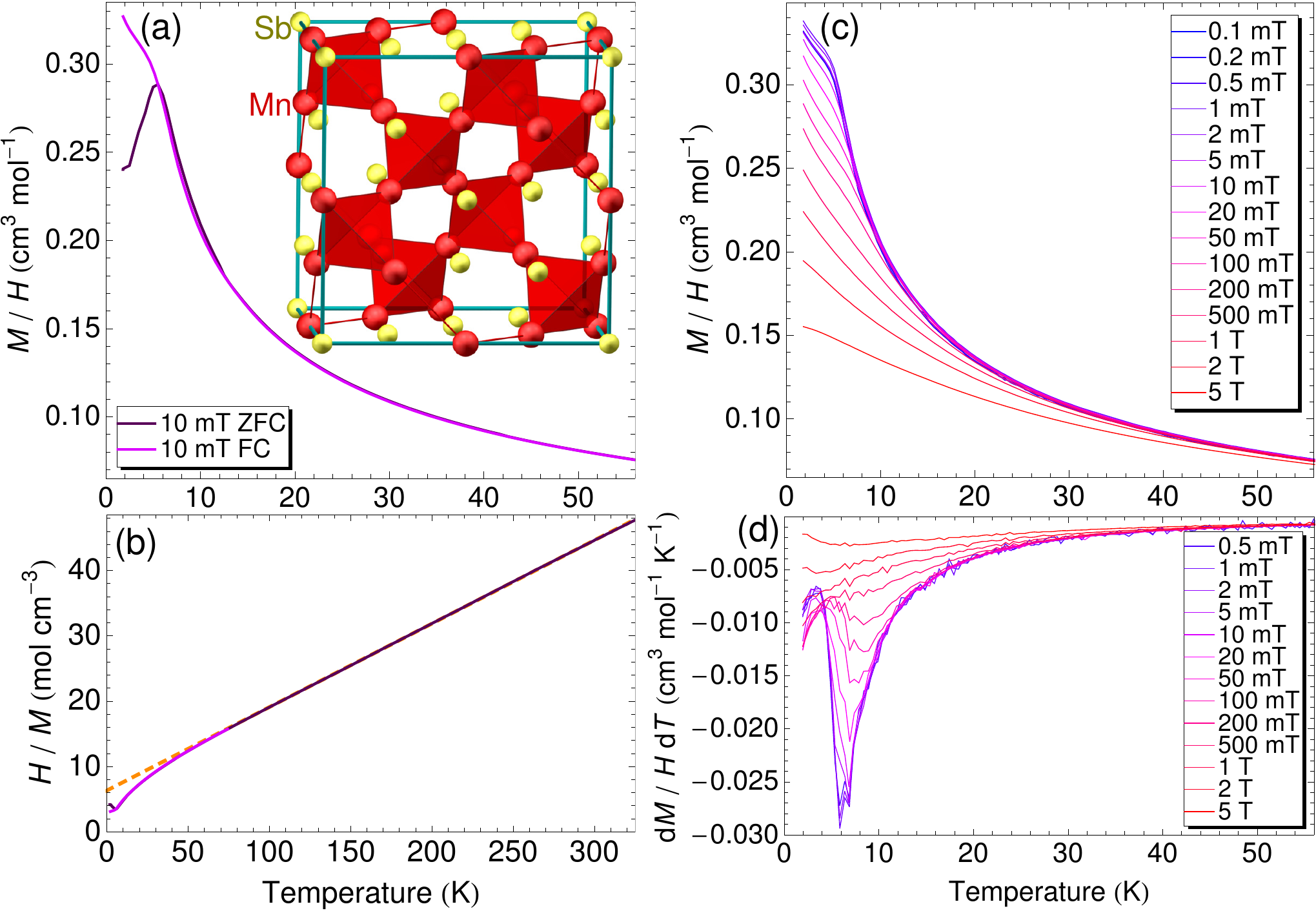}
  \caption{\label{fig:MvsT}Magnetization $M/H$ of \py.  (a)
    Zero-field-cooled (ZFC) and field-cooled (FC) magnetization in
    10\,mT, showing a transition at 5.5\,K (b) Inverse magnetization,
    showing a deviation from Curie-Weiss behavior (dashed line) below
    $\sim$80\,K.  (c) Field-dependence of the FC magnetization.  (d)
    Temperature derivative of the data in (c).  The crystal structure
    is depicted in the inset to (a), with Mn tetrahedra in red and the
    Sb sublattice in yellow.}
\end{figure}

When prepared by conventional solid-state synthesis, \Mn\ forms in a
distorted, chiral variant of the trigonal Weberite
structure\cite{Scott1987,Chelazzi2013,Peets-P2}, but the desired
pyrochlore polymorph (Fig.~\ref{fig:MvsT}a inset) can be stabilized at
lower temperatures\cite{Brisse1972,Zhou2008,Zhou2010}.  We prepared
powder samples of \py\ following Brisse\cite{Brisse1972}, with
precursor `antimonic acid' prepared from SrCl$_5$ (Alfa Aesar,
99.997\%) and deionized water as in Ref.~\onlinecite{Stewart1970}.
The precursor was ground with Mn(Ac)$_2\cdot$4H$_2$O (Aldrich,
99.99\%) then reacted for 12h at a sequence of temperatures from 50 to
550$^\circ$C in Al$_2$O$_3$ crucibles in air.  Powder diffraction and
magnetic measurements were used to verify phase purity and Mn site
occupancy.

X-ray powder diffraction patterns were collected using a Bruker D8
Discover diffractometer with a Cu$K\alpha$ source, while a Rigaku
Miniflex II was used to collect patterns on precursor `antimonic acid'
and monitor reaction completeness.  Magnetization measurements were
performed in a Quantum Design MPMS-XL magnetometer in RSO measurement
mode: a $\sim$10\,mg powder sample was gently compressed inside a
gelatin capsule, which was closed with Kapton tape and inserted into a
plastic straw.  The sample holder contribution was below the noise
level.  AC susceptibility was measured in the same magnetometer using
the DC sample transport, at zero applied field.  Specific heat was
measured on a thin plate of pressed powder by the relaxation time
method in fields up to 9\,T in a Quantum Design PPMS.  As an
approximate phonon baseline, trigonal Weberite Sr$_2$Sb$_2$O$_7$
ceramic prepared by standard solid state synthesis was used.  Powder
neutron diffraction was performed at the ECHIDNA diffractometer at the
OPAL research reactor at ANSTO, Australia, from 6.5 to 164$^\circ$ in
steps of 0.05$^\circ$, with neutron wavelengths of 1.6220\,\AA\ (room
temperature) and 2.4395\,\AA\ (low temperature).  Diffraction data
were Rietveld-refined in FullProf by the least-squares
method\cite{FullProf}.  Throughout this paper, molar quantities refer
to formula units.

The temperature-dependent DC magnetization of \py, shown in
Fig.~\ref{fig:MvsT}, indicates a magnetic transition at $\TSG=5.5$\,K,
below which the zero-field-cooled and field-cooled magnetization data
(Fig.~\ref{fig:MvsT}a) diverge, implying that the spins order or
freeze.  This transition broadens and moves to higher temperature in
field (Fig.~\ref{fig:MvsT}c), becoming indistinct above $\sim$0.5\,T.
This is more clearly visible in the derivative (Fig.~\ref{fig:MvsT}d),
which also shows that there is no evidence for any other transition.
In particular, the transition reported around
41\,K\cite{Zhou2008,Zhou2010} is absent.  As shown in
Fig.~\ref{fig:MvsT}b, above $\sim$80\,K the magnetization obeys the
Curie-Weiss law with a Curie-Weiss temperature $\TCW=-49$\,K
indicating predominantly antiferromagnetic interactions.  Deviations
from Curie-Weiss behaviour below $\sim$80\,K indicate the onset of
short-range correlations more than an order of magnitude above \TSG,
while the approximately order-of-magnitude difference between
\TCW\ and \TSG\ indicates strong frustration.  The paramagnetic moment
of 5.59\,$\mu_B$ is consistent with the spin-only expectation of
5.92\,$\mu_B$ for high-spin $3d^5$, confirming the absence of an
orbital contribution and demonstrating that \py\ is indeed a
Heisenberg spin system.

\begin{figure}[htb]
  \includegraphics[width=\columnwidth]{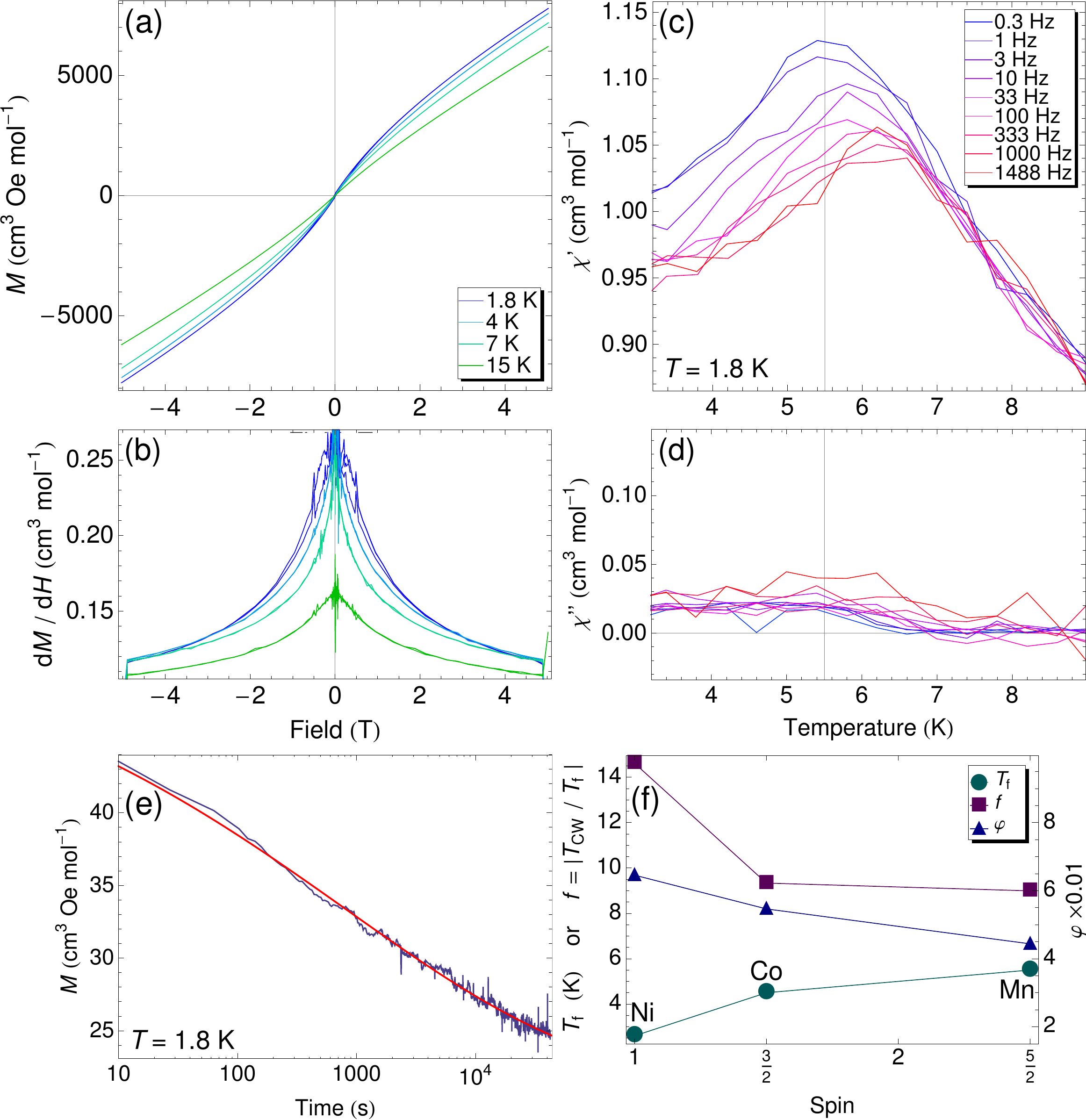}
  \caption{\label{fig:MH}Magnetic response of \py.  (a) $M(H)$ loops;
    slight hysteresis at 1.8 and 4\,K is more evident in (b) the
    derivative $dM/dH$.  This smooth field and temperature evolution
    shows no evidence for an abrupt phase transition.  Jumps in the
    derivative correspond to changes in sweep rate.  (c) A broad peak
    in the real part of the zero-field AC susceptibility at
    $\TSG=5.5$\,K, the same temperature identified in $M(T)$, shifts
    substantially as the drive frequency increases. (d) The loss,
    $\chi''$, is weakly positive below the transition.  (e) Strong
    relaxation is visible at 1.8\,K.  The sample was cooled in a field
    of 0.1\,T, then the field was removed and the zero-field
    relaxation monitored.  (f) Evolution of the intrinsic transitions
    and frustration in $M_2$Sb$_2$O$_7$ for $M$=Ni, Co, Mn, based on
    Ref.~\onlinecite{Zhou2010}.}
\end{figure}

The field dependence of the magnetization (Fig.~\ref{fig:MH}a) is
S-shaped and exhibits hysteresis at low temperatures, which persists
to at least 1.5\,T with no clear onset field.  The derivative
(Fig.~\ref{fig:MH}b) shows clearer evidence of the hysteresis and a
lack of any sharp phase transition.  Abrupt changes in the hysteresis
loops where the field sweep rate was changed indicate strong
relaxation effects on the timescale of minutes.  To verify this,
Fig.~\ref{fig:MH}e shows the results of field-training the sample by
cooling to 1.8\,K in a 0.1\,T field, then monitoring the ensuing
magnetization decay after reducing the field to zero.  The
magnetization is well described by a stretched exponential $M(t) = M_0
+ M_R\exp(-(t/t_P)^{1-n})$\cite{Mydosh1986,Binder1986}, with a
non-relaxing offset $M_0 = 21.6(2)$\,cm$^3$\,Oe\,mol$^{-1}$, a
substantial relaxing component $M_R =
31.7(14)$\,cm$^3$\,Oe\,mol$^{-1}$, time scale $t_P$ of 835(112)\,s,
and exponent $n$ of 0.7843(97).  An S-shaped hysteresis loop with
rate- or time-dependent hysteresis is one key hallmark of a spin
glass\cite{Mydosh1986}.

The zero-field AC susceptibility $\chi$ measured with an excitation
amplitude of 1\,Oe, shown in Figs.~\ref{fig:MH}(c,d), also shows
fingerprints of strong relaxation.  A broad peak at low frequencies in
the real component $\chi'$ agrees well with the magnetization
transition, then shifts to significantly higher temperatures as the
frequency increases.  The imaginary component, $\chi''$, becomes
weakly positive on cooling through the transition, indicating
dissipation as the low-frequency measurement field aids the system in
finding a ground state.  By tracking the frequency dependence of the
intrinsic low-temperature transition, the Mydosh parameter $\varphi =
\Delta T_N / T_N\Delta\log f = 0.044$ can be extracted.  This is
comparable to 0.053 and 0.064 found for pyrochlore Co$_2$Sb$_2$O$_7$
and Ni$_2$Sb$_2$O$_7$, respectively\cite{Zhou2010}, which undergo
similar transitions into their low-temperature ground states.  The
three compounds' progression in transition temperatures, frustration
factors $f\equiv \left\vert \TCW/\TSG\right\vert$ and Mydosh
parameters is plotted in Fig.~\ref{fig:MH}f, based on
Ref.~\onlinecite{Zhou2010}.  Given that the Co and Ni analogs have
unquenched orbital moments, the evolution is surprisingly mundane.

\begin{figure}[htb]
  \includegraphics[width=\columnwidth]{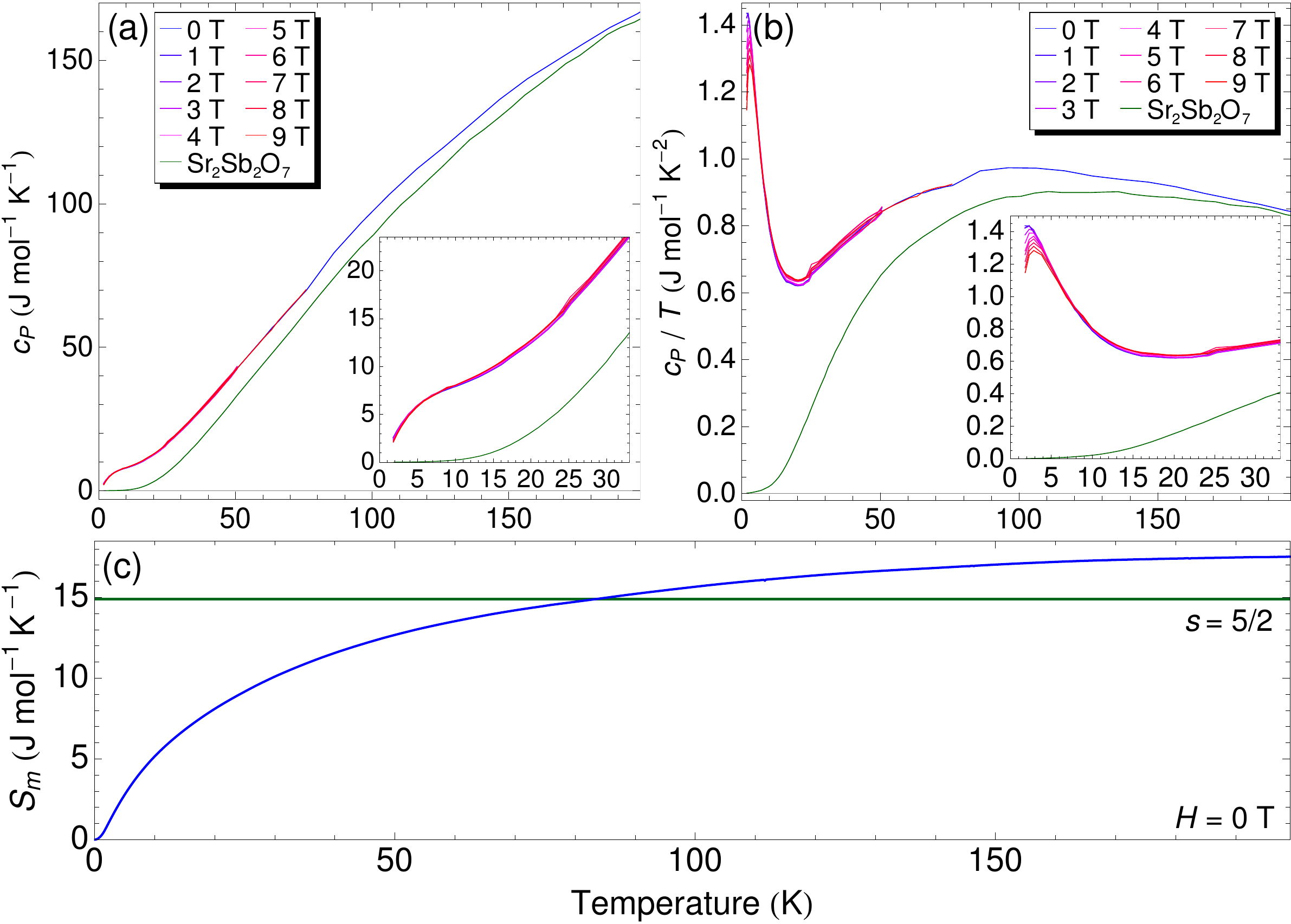}
  \caption{\label{fig:cP}Specific heat of \py.  (a) There is a hump in
    $c_P(T)$ at low temperatures, and no feature around 41\,K.
    Low-temperature data are highlighted in the inset, and data on
    Weberite Sr$_2$Sb$_2$O$_7$ serve as a baseline.  (b) The broad
    peak in $c_P(T)/T$ is modified only slightly in field.  Magnetic
    entropy is apparent to at least 40\,K.  (c) The approximate
    magnetic entropy calculated by subtracting the Sr$_2$Sb$_2$O$_7$
    data. The entropy expected for $s=5/2$ cannot all be found below
    $\sim$20\,K in the $c_P/T$ peak.}
\end{figure}

The specific heat in Fig.~\ref{fig:cP} shows a large and very broad
hump of magnetic entropy at low temperatures, centered around 2.5\,K,
roughly half the temperature found {\it via} magnetization.  It does
not resemble a usual first- or second-order phase transition, and we
have not attempted to extract any more precise estimate for the
transition temperature.  In applied magnetic fields the hump moves to
higher temperatures as does the transition in the magnetization, but
it also appears to narrow slightly.  Such behavior is readily
explained by magnetic field relieving the frustration, and is common
in spin glasses.  Specific heat data for
Sr$_2$Sb$_2$O$_7$\cite{Knop1980}, a nonmagnetic insulator forming in
the related trigonal Weberite structure\cite{Verscharen1978,Cai2009},
are included for comparison.  However, the phonons in \py\ clearly
freeze out at significantly lower temperatures, making any subtraction
of the phonon background inexact.  A very approximate measure of the
magnetic entropy can be extracted by subtracting the Sr$_2$Sb$_2$O$_7$
data as a phonon background, but the only conclusion to be drawn here
is that most of the entropy expected from $s=5/2$ $3d^5$ is not
released near the transition.


\begin{figure}[htb]
  \includegraphics[width=\columnwidth]{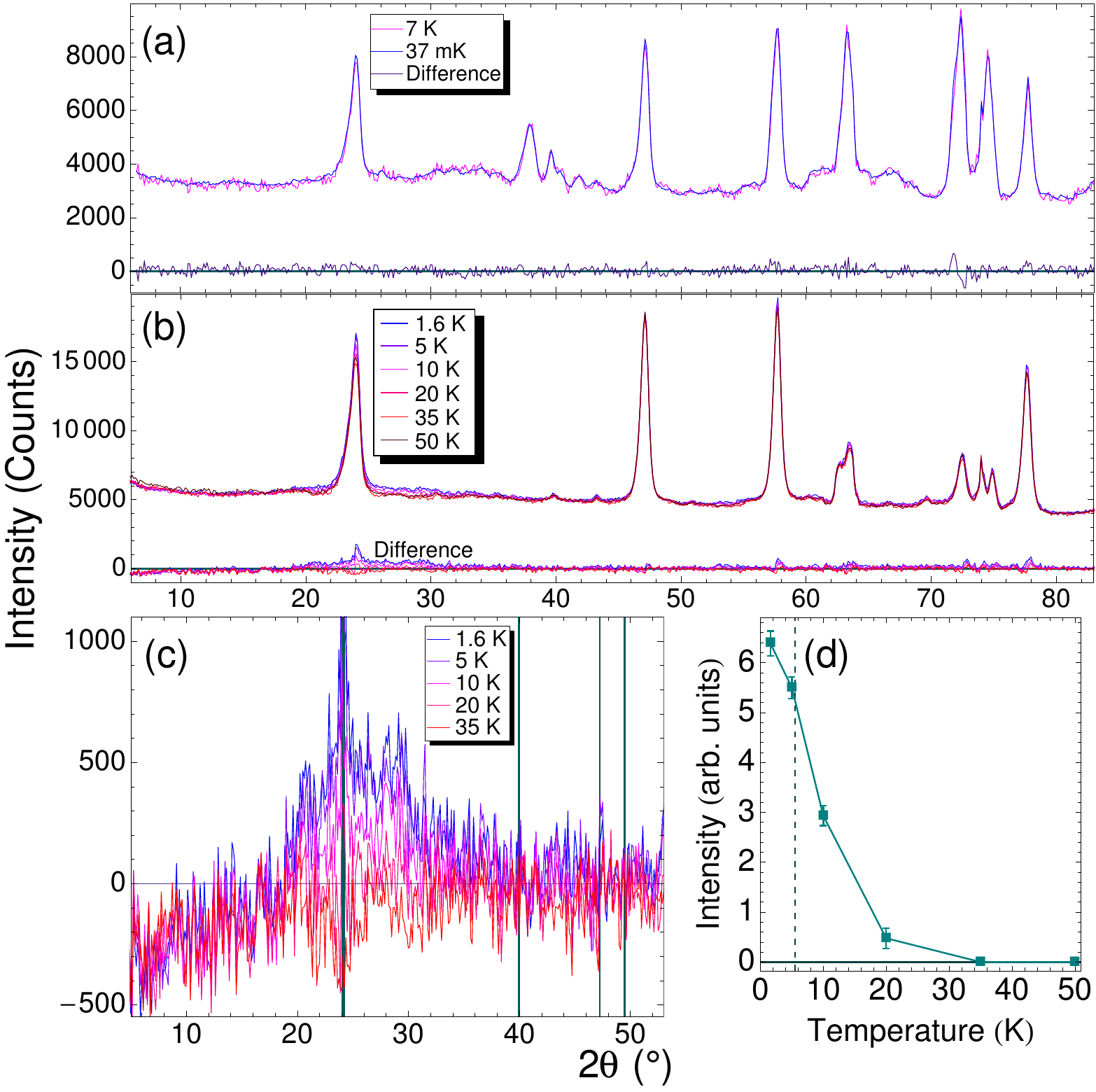}
  \caption{\label{neutron}Low-temperature neutron diffraction on \py.
    (a, b) Diffraction patterns at low temperature, through \TSG,
    demonstrate the absence of any new magnetic or structural peaks.
    Differences between 7\,K and 37\,mK data (a) and from the 50\,K
    data (b) are shown.  Several peaks originate from the sample
    holder --- Cu in (a), Al in (b).  (c) Difference from data taken
    at 50\,K, highlighting the diffuse contribution. Nuclear Bragg
    positions are marked. (d) Temperature dependence of the diffuse
    peak.}
\end{figure}

Low-temperature neutron diffraction patterns collected through
\TSG\ are shown in Figs.~\ref{neutron}a and \ref{neutron}b.  The
transition introduces no new peaks nor significant changes to the
intensities of existing peaks, as further demonstrated by taking the
difference across the transition.  It is worth noting that there is no
low-temperature structural transition that could reduce the
frustration.  Fig.~\ref{neutron}c highlights the presence of a diffuse
magnetic component near the (111) nuclear Bragg peak at a $d$-spacing
of $\sim$5.8\,\AA, which becomes indistinct above 20\,K, in line with
the magnetization and specific heat results.  The temperature
dependence is shown in Fig.~\ref{neutron}d, based on fitting each
difference pattern to a Gaussian with a sloping background.  \py\ now
meets all the canonical criteria and has all the earmarks of a spin
glass: the ZFC and FC magnetization differ, strong relaxation is
observed, the AC susceptibility is strongly frequency dependent, there
is only a broad hump below the transition in the specific heat, and
long-range order is absent.

\begin{figure}[htb]
  \includegraphics[width=\columnwidth]{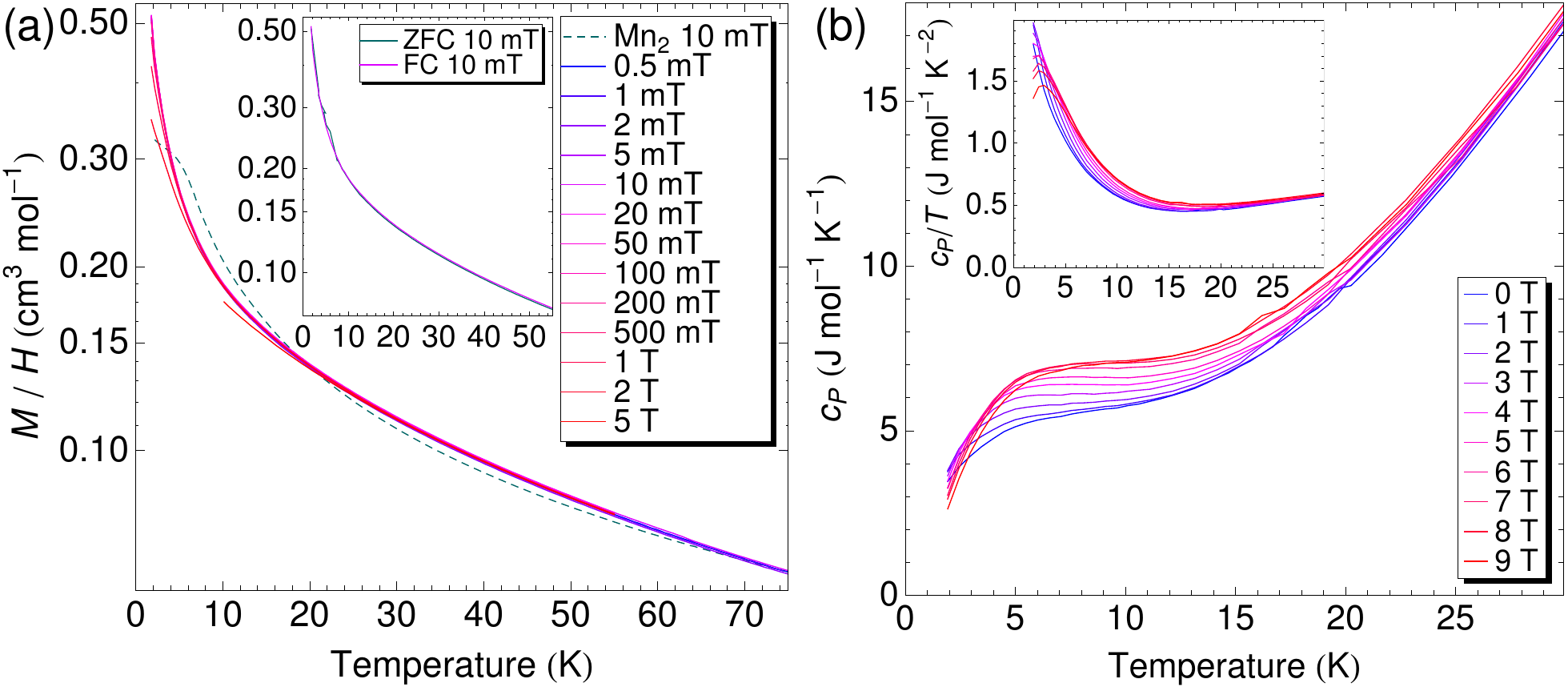}
  \caption{\label{deficient}Properties of Mn-deficient \py.  (a) There
    is no transition in the field-cooled or zero-field-cooled (inset)
    magnetization.  (b) Magnetic fields shift the magnetic entropy to
    higher temperature.}
\end{figure}



Mn ions in \py\ populate an extremely high-symmetry lattice which
should be essentially free from chemical disorder, so there should be
nothing extrinsic to the spin system that could locally lock the spins
and force a glass transition.  Since \py\ most likely forms through
substituting Mn atoms into the pre-existing pyrochlore matrix of
Sb$_2$O$_5\cdot n$H$_2$O\cite{Brisse1972}, a manganese deficiency is
possible and manganese vacancies may be the dominant defect.  X-ray
diffraction indicated Mn deficiencies in incompletely-reacted material
and those prepared from Mn-poor mixtures, but the Mn site refined to
full occupancy in all samples reported above.  Mn-deficient samples
show a significant suppression of \TSG, often with no evidence for
even the onset of a transition above 1.8\,K, but with much stronger
field-enhancement in the specific heat --- an incompletely-reacted
example is shown in Fig.~\ref{deficient}.  This suppression by
disorder strongly suggests that the magnetic freezing is intrinsic.
As thermal fluctuations diminish on cooling, it presumably becomes
more difficult to rotate any one spin because of the increasing number
of spins coupled to it: effectively a diverging mass term.
Intriguingly, this behavior is not observed in FeF$_3$, which enters
an all-in-all-out magnetic ground state\cite{Ferey1986,Reimers1992}.
The latter has been explained through density functional
theory\cite{Sadeghi2015}, but similar calculations for \py\ have not
been performed.  In FeF$_3$ nearest-neighbor antiferromagnetic
Heisenberg, biquadratic, and Dzyaloshinskii-Moriya interactions were
all required to adequately model the spin system.  In \py, which has a
smaller unit cell and rather different ligand coordination, these may
be expected to have completely different strengths.  In particular,
the Mn tetrahedron in \py\ contains a central O2 atom, offering a much
more direct exhange pathway.  Modelling the interactions in this
material and the resulting magnetic ground state will be an important
avenue for future theoretical exploration.

One important question is why we observe a bulk ordering transition at
5.5\,K while previous reports placed it around
41\,K\cite{Zhou2008,Zhou2010}.  We attribute the previous results to
trace quantities of unreacted Mn$_3$O$_4$, which has a ferrimagnetic
transition around this temperature\cite{Dwight1960,Jensen1974}.
Samples which were Mn-rich or incompletely reacted often exhibited a
strong transition around 41\,K in the magnetization even when
phase-pure by x-ray diffraction; however, there was no trace of the
transition in the specific heat.  Because the Mn$_3$O$_4$
ferrimagnetic moments are large and readily field-trained, even trace
quantities can dominate the low-field magnetization.  This makes it
crucial to completely eliminate this impurity if the intrinsic physics
of \py\ are to be investigated.  Previous measurements on \py\ display
hints of intrinsic behaviour at low temperatures in the inverse
magnetization\cite{Zhou2008}, but the data otherwise concentrate,
understandably, on the obvious 41\,K transition.

In summary, the $3d$-electron Heisenberg pyrochlore \py\ undergoes a
transition around 5.5\,K into a spin glass state with dominantly
antiferromagnetic interactions.  Short-range magnetic correlations
survive to much higher temperatures, and strong frustration is
evident.  The manganese content can be tuned continuously from zero to
full occupancy, offering a very interesting control knob for future
disorder studies.  Measurements to lower temperatures will be required
to establish whether the specific heat resembles that in kagome
order-by-disorder systems\cite{Ramirez1990,Chalker1992,Ritchey1993}.
The physical properties of \py\ bear a striking similarity to those of
the magnetic pyrochlores Ni$_2$Sb$_2$O$_7$ and
Co$_2$Sb$_2$O$_7$\cite{Zhou2010}, and more recently NaSrCo$_2$F$_7$,
NaCaNi$_2$F$_7$, NaCaFe$_2$F$_7$, and
NaSrFe$_2$F$_7$\cite{Krizan2015,Krizan2015b,Sanders2016}, which should
all have unquenched orbital moments.  This suggests that spin-orbit
coupling may be so weak in $3d$ transition metal pyrochlores as to be
effectively irrelevant to the frustrated spins, greatly broadening the
suite of materials available for investigating effective Heisenberg
physics on the most frustrated known three-dimensional lattice.  The
very recently reported NaSrMn$_2$F$_7$\cite{Sanders2016}, with
quenched orbital moments but A-site cation disorder, also behaves in a
similar manner.

\begin{acknowledgments}
This work was supported by the Institute for Basic Science (IBS) in
Korea (IBS-R009-G1).  The authors are indebted to M.\ Gingras for
stimulating discussions.  We acknowledge the support of the Bragg
Institute, Australian Nuclear Science and Technology Organisation, in
providing neutron research facilities used in this work.
\end{acknowledgments}

\bibliography{Mn2Sb2O7}

\appendix
\section{Supplementary Information:  Structure Refinement}
\input{supp_content}

\end{document}

%% file: supp_content.tex
\begin{figure}[hb!]
  \includegraphics[width=\columnwidth]{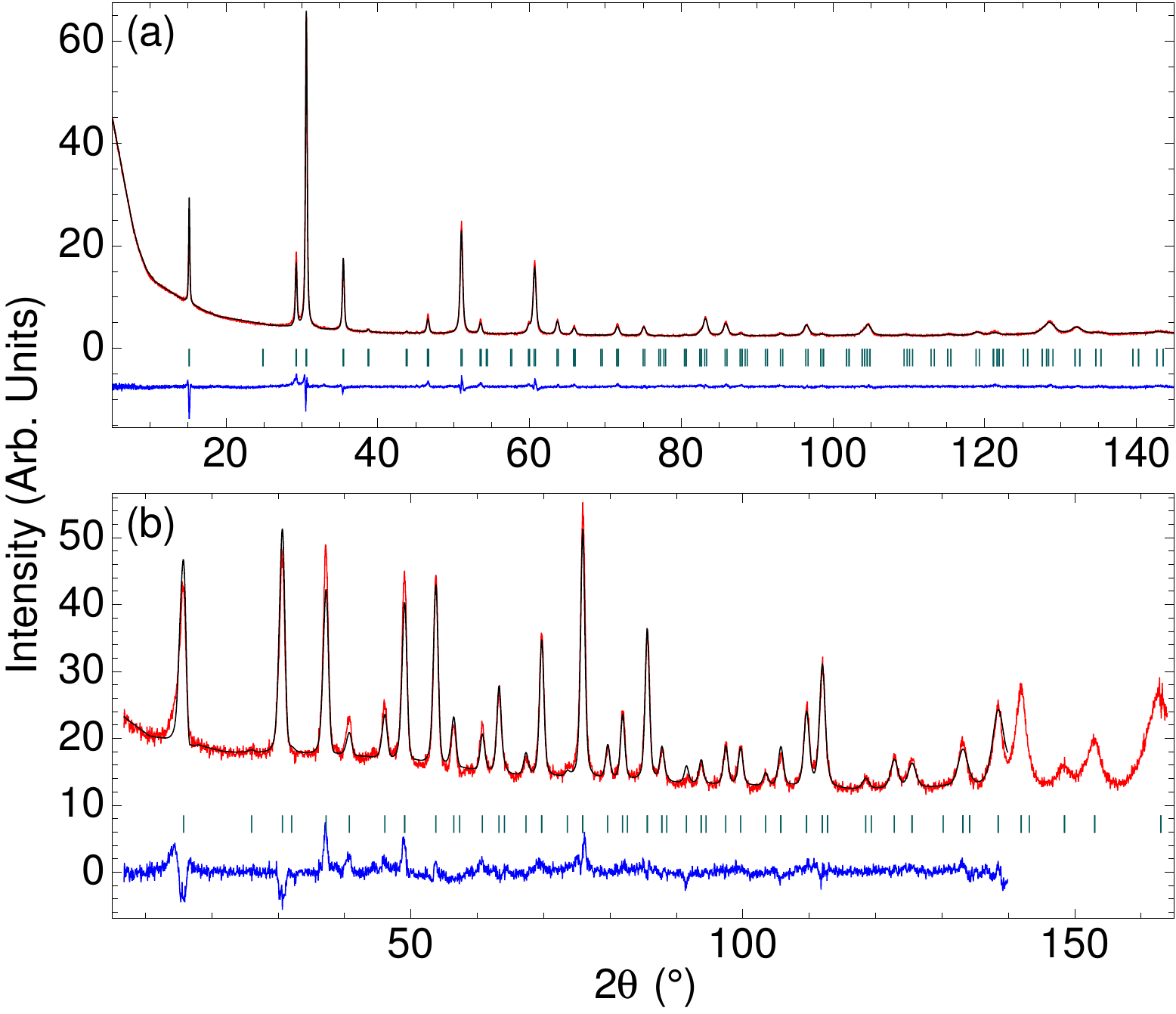}
  \caption{\label{fig:xrd}Refined room temperature powder
    diffractograms for \py:  (a) x-ray and (b) neutron diffraction.
    Data are in red, the fit is in black, the residual is in blue, and
    vertical bars mark the calculated Bragg positions.  The residual
    has been shifted for clarity.}
\end{figure}

Fig.\ \ref{fig:xrd} shows a joint refinement of x-ray and neutron
powder diffraction data on \py\ at room temperature, in the
$Fd\overline{3}m$ space group (\#\,227).  The lattice parameter is
$a=10.1073(4)$\,\AA\ in the neutron data and 10.1205(11)\,\AA\ in the
x-ray data.  The overall $R$-factors based on all points and not
corrected for background were $R_p=3.89$\%\ and $R_{wp}=5.00$\%\ for
the neutron pattern and $R_p=2.98$\%\ and $R_{wp}=4.37$\%\ for the
x-ray pattern, with a global $\chi^2$ of 0.757, based on 67 neutron
reflections and 140/2 reflections for the x-ray pattern.

\begin{table}[htb!]
  \caption{\label{tab:pyr}Refined atomic positions for \py\ in space
    group $Fd\overline{3}m$ (\#\,227) at room temperature.}
  \begin{tabular}{lcllll}\hline\hline
    Site & Mult.& $x$ & $y$ & $z$ & $B_{iso}$ (\AA$^2$)\\ \hline
    Mn & $16d$ & 0.5 & 0.5 & 0.5 & 2.85(16)\\
    Sb & $16c$ & 0.0 & 0.0 & 0.0 & 2.11(10)\\
    O1 & $48f$ & 0.32744(16) & 0.125 & 0.125 & 0.604(22)\\
    O2 & $8b$ & 0.375 & 0.375 & 0.375 & 0.604(22)\\ \hline\hline
  \end{tabular}
\end{table}

The high background in the neutron data is attributed to the presence
of hydrogen --- the material is prepared at relatively low
temperatures from a hydrate and an organic Mn source, both of which
are rich in H atoms, and takes the form of an extremely fine powder
with a high surface-to-volume ratio and considerable potential to host
adsorbed water.  In analyzing this sample, the Mn site was allowed to
deviate from full occupancy as a test for Mn vacancies.  The x-ray
refinement was not improved --- the refined Mn content was 1.99(5),
indicating full occupancy.  However, the joint refinement suggested
vacancies on the Mn and Sb sites when the occupancy of each was
allowed to vary, and excess scattering on the O1 site in particular.
This may indicate the presence of remnant hydrogen atoms, but without
a model for what species would be present ({\it e.g}., H$_2$O,
H$_3$O$^+$), we were not able to reach firm conclusions.  The refined
structure of \py\ is summarized in Tab.\ \ref{tab:pyr}, and selected
bond lengths and bond angles that may be of particular interest for
the magnetic interactions are presented in Tab.\ \ref{tab:bond}.

\begin{table}[htb!]
  \caption{\label{tab:bond}Selected bond lengths and bond angles for
    \py\ at room temperature, based on the refinement presented in
    Fig.~\ref{fig:xrd} and Tab.~\ref{tab:pyr}.}
  \begin{tabular}{lr@{.}lr@{.}l}\hline\hline
    Sites & \multicolumn{2}{c}{Bond length} & \multicolumn{2}{c}{Bond angle} \\ \hline
    Mn--O1 & 2&5001(8)\,\AA & \multicolumn{2}{c}{}\\
    Mn--O2 & ~2&19115(14)\,\AA & \multicolumn{2}{c}{}\\
    Sb--O1 & 1&9532(7)\,\AA & \multicolumn{2}{c}{}\\
    Mn--O1--Mn & \multicolumn{2}{c}{} & 91&38(4)$^\circ$\\
    Mn--Mn & 3&5781(3)\,\AA & \multicolumn{2}{c}{}\\
\hline\hline
  \end{tabular}
\end{table}